\documentclass[]{article}

\begin{document}

\title{Charmonium and Bottomonium from Classical $SU(3)$ Gauge Configurations}

\author{R. A. Coimbra, O. Oliveira \\
        Centro de F\'{\i}sica Computacional, Universidade de Coimbra, \\
           3004-516 Coimbra, Portugal}

\maketitle

\begin{abstract}
The charmonium and bottomonium spectra computed from a potential defined from
a single gauge configuration, obtained from solving the classical field
equations, is discussed. The theoretical spectra shows good agreement with
the measured states.

A discussion of possible interpretations, within the same non-relativistic
potential model, for the new charmonia states
$X(3872)$, $\chi_{c1} (2P)$ and $Y(4260)$ is performed. In particular, we
give predictions for electromagnetic E1 transitions for various scenarios.
\end{abstract}


\section{Introduction and Motivation}

In \cite{OlCo06} it was proposed a generalized Cho-Faddeev-Niemi ansatz for
the SU(3) gauge fields. For the simplest form of the ansatz, the classical 
field equations were solved and a potential for heavy quarkonia motivated. 
The single configuration potential is coulombic for short interquark distances
and grows exponentially for large interquark distances. In this way,
the potential provides quark confinement. Then, assuming that the quark 
interaction is a pure vectorial interaction, the spectra of charmonium was 
investigated. 

As described in \cite{OlCo06}, the single configuration potential is able to 
describe the charmonium spectra with an error of less than 3\% 
($\sim 100$ MeV). However, the prediction for the $1S$ and $2S$
hyperfine splitting is about half of the experimental value. This result is
probably due to the definition of the potential from a single gauge 
configuration. In the same work, the theoretical predictions for the leptonic 
widths of $1^{--}$ states where computed and the results are in-line with the 
experimental values.

In \cite{CoOl06} the investigation was extended to include the E1 
electromagnetic transitions together with an analysis of the bottomonium
system. For $b \overline b$, the spectra is reproduced with an error of less
than 1\% ($\sim 100$ MeV) and the theoretical prediction for the 
$b \overline b$ leptonic widths reproduces the same level of accurary as found
in charmonium. 

In what concerns the electromagnetic transitions, the single configuration 
potential is able to reproduce well the quoted particle data book numbers 
\cite{pdg}.

\section{Charmonium and Bottomonium Spectra}

In table \ref{spectra} we show the theoretical $J^{PC} = 1^{--}$ spectra; see
\cite{OlCo06,CoOl06} for details.
The third column is the deviation of the theoretical mass to the 
experimental measured mass, in MeV. The overall agreement between theoretical
and particle spectra is good. The only particles which don't fit well in the
theoretical spectra are $\psi (4040)$ and $\psi (4415)$. In what concerns these
two states, the experimental information is scarce and the particle data
book comments that the ``interpretation of these states as a single
resonance is unclear because of the expectation of substantial threshold 
effects in this energy region''. Curiously, both particle masses are 
essentially the sum of $J / \psi$ with light $J^{PC} = 0^{++}$ mesons.

\begin{table}
\begin{center}
\begin{tabular}{lrrr|lrrr}
\hline
  & \multicolumn{3}{r|}{\textbf{Charmonium}}
  & \multicolumn{4}{r}{\textbf{Bottomonium}} \\
\hline
 $1^3S_1$ &  3097 &    0 & $J / \psi (1S)$ &  
 $1^3S_1$ &  9460 &    0 & $\Upsilon (1S)$ \\
 $2^3S_1$ &  3659 &  -27 & $\psi (2S)$ &  
 $2^3S_1$ & 10023 &    0 & $\Upsilon (2S)$ \\
 $1^3D_1$ &  3688 &  -83 & $\psi (3770)$ &  
 $1^3D_1$ & 10159 &    - &                  \\
 $3^3S_1$ &  4164 &  -95 & $Y (4260)$ &  
 $3^3S_1$ & 10385 &   30 & $\Upsilon (3S)$  \\
 $2^3D_1$ &  4155 &    2 & $\psi (4160)$ &
 $2^3D_1$ & 10476 &    - &                  \\
 $4^3S_1$ &  4669 &    - &   &
 $4^3S_1$ & 10727 &  148 & $\Upsilon (4S)$ \\
 $3^3D_1$ &  4636 &    - & &
 $3^3D_1$ & 10796 &  -69 & $\Upsilon (10860)$ \\
          &       &      & &
 $5^3S_1$ & 11065 &   46 & $\Upsilon (11020)$ \\
\hline
\end{tabular}
\end{center}
\caption{$J^{PC} = 1^{--}$ charmonium and bottomnium spectra. The table shows
the theoretical state, the mass prediction, the difference between the 
experimental measured mass and the theoretical prediction. All numbers are 
in MeV.}
\label{spectra}
\end{table}

The work reported in \cite{OlCo06,CoOl06} and summarized in table 
\ref{spectra} is based on the nonrelativistic analysis of a confining 
potential derived from a single configuration. Our
previous work suggests that either one should include the contribution from 
other configurations and/or one should perform a coupled-channel analysis. 
We are currently engaged in extending our previous studies in both ways. 
Anyway, in what concerns the spectra, the single channel analysis shows that 
the potential is able to explain the observed states if one allows for an error
of $\sim 100$ MeV. 

In the following we report on the predictions of the single channel 
analysis for the potential obtained in \cite{OlCo06} for the new charmonium 
states $X(3872)$, $\chi_{c2} (2P)$ and $Y(4260)$ (we follow the particle data
book notation). In order to be able to distinguish the possible quantum number
assignements, when possible, we also report on our predictions for 
electromagnetic E1 transitions. We call the reader attention to the good
agreement between theoretical predictions and experimental measures of
charmonium and bottomonium electromagnetic widths - see tables 4, 5,6 of 
\cite{CoOl06}.

\section{$X(3872)$}

In what concerns the quantum numbers of $X(3872)$, experimentaly only the 
parity, $C = +$, is known. Belle Collaboration 
\cite{Belle} has performed an analysis of possible quantum numbers
and conclude in favor of $J^{PC} = 1^{++}, 2^{++}$. 
In table \ref{X3872}, we report the charmonium states compatible with these
quantum numbers and whose mass differs from the $X(3872)$ by $100$ MeV, 
including the E1 electromagnetic widths for the assignement favoured by 
Belle data.

\begin{table}
\begin{center}
\begin{tabular}{lrrlr}
\hline
  & \multicolumn{1}{r}{Mass}
  & \multicolumn{1}{r}{$J^{PC}$}
  & \multicolumn{2}{r}{E1 Electromagnetic Transition} \\
  & \multicolumn{1}{r}{(MeV)}
  & \multicolumn{1}{r}{}
  & \multicolumn{2}{r}{(KeV)} \\
\hline
$2^3P_1$ & 3938 & $1^{++}$ & $\longrightarrow \psi (2S) + \gamma$     &  63 \\
         &      &          & $\longrightarrow J / \psi (1S) + \gamma$ &  48 \\
$1^3F_2$ & 3932 & $2^{++}$ & &   \\
\hline
\end{tabular}
\end{center}
\caption{$X(3872)$ possible interpretations and E1 electromagnetic
transitions.}
\label{X3872}
\end{table}
%
%

\section{$\chi_{c2} (2P)$}

According to the particle data book, this is a $J^{PC} = 2^{++}$ with a mass
of $3929 \pm 5$ MeV. In our model, the $J = 2$ states around this mass 
value are $2^3P_2$, with $J^{PC} = 2^{++}$ and mass 4048 MeV; the
$1^3F_2$, with $J^{PC} = 2^{++}$ and mass 3932 MeV.

If this state is a  $2^3P_2$ state, it has a large E1 width of 140 KeV for
the transition to $\psi ' (2S)$ and a E1 width of 59 KeV to $J / \psi (1S)$.

\section{$Y(4260)$}

This $c \overline c$ state is a $J^{PC} = 1^{--}$ with a mass of 
$4259^{+8}_{-10}$ MeV. According to the particle data book, the 
``interpretation as due to two interfering resonances is not excluded''.
In our study, possible candidates with a mass between $\sim 4160$ MeV and
$\sim 4360$ MeV are: $3^3S_1$, 4155 MeV; $2^3D_2$, 4327 MeV; $2^3D_2$, 
4230 MeV; $2^1D_2$, 4230 MeV; $1^3G_4$, 4260; $1^3G_3$, 4161 MeV; 
$3^3P_0$, 4300 MeV. Only the state $3^3S_1$ has $J^{PC} = 1^{--}$, as it 
should be for a state producted via initial state radiation.

If $Y(4260)$ is a $3^3S_1$ state, its larger E1 electromagnetic
widths are transitions to $\chi (2P)$ states, namely: 
$2^3P_0$, $\Gamma = 358$ KeV; $2^3P_1$, $\Gamma = 475$ KeV; 
$2^3P_2$, $\Gamma = 233$ KeV. The corresponding widths for $\chi (1P)$ states
being: $1^3P_0$, $\Gamma = 4$ KeV; $1^3P_1$, $\Gamma = 9$ KeV; 
$1^3P_2$, $\Gamma = 13$ KeV, making them hard to measure. Given the values
for the various widths, it seems that a combine investigation of $Y(4260)$ 
and $\chi_c (2P)$ could be helpfull in understanding the nature of this 
particle.


\section*{Acknowledgments}
R. A. C. acknowldges F.C.T. for financial support, grant SFRH/BD/8736/2002.
This work was partly supported by F.C.T. under contract POCI/FP/63436/2005.



\begin{thebibliography}{9}
\bibitem{OlCo06}
O. Oliveira, R. A. Coimbra, hep-ph/0603046.

\bibitem{CoOl06}
R. A. Coimbra, O. Oliveira, hep-ph/0610142.

\bibitem{pdg}
W.-M. Yao \textit{el al.} (Particle Data Book), \emph{J. Phys.}
\textbf{G33}, 1 (2006).

\bibitem{Belle}
K. Abe \textit{et al}. (Belle Collaboration), hep-ex/0505038.

\end{thebibliography}
\end{document}